\documentclass[10pt,conference]{IEEEtran}
\IEEEoverridecommandlockouts

\usepackage[utf8]{inputenc}
\usepackage[english]{babel}
\usepackage{fontenc}
\usepackage[backend=biber,style=numeric,sorting=none]{biblatex}
\usepackage{amsmath, amssymb}
\usepackage{hyperref}
\usepackage{algorithm}
\usepackage{algpseudocode}
\usepackage{csquotes}
\usepackage{amsthm}
\usepackage{array}
\usepackage{booktabs}
\usepackage{xcolor}
\usepackage{graphicx}
\usepackage{orcidlink}
\usepackage{tikz}
\graphicspath{{figs/}}
\usetikzlibrary{positioning,backgrounds,fit,graphs,matrix,patterns}
\usetikzlibrary{external}
\usetikzlibrary{shapes.multipart,arrows.meta}

\newcommand{\ket}[1]{\lvert #1 \rangle}
\newcommand{\comm}[2]{\left[#1, #2\right]}
\newcommand{\tr}{\text{tr}}
\newcommand{\G}{\mathcal{G}}

\newcommand{\So}{\mathfrak{so}}
\newcommand{\Su}{\mathfrak{su}}
\newcommand{\Sp}{\mathfrak{sp}}

\newcommand{\Continue}{\textbf{continue}}

\newtheorem{definition}{Definition}

\usepackage{xcolor}

\definecolor{unlitgeclr}{HTML}{D83C3C}
\definecolor{litgeclr}{HTML}{0C7BDC}
\definecolor{unlitclr}{HTML}{1E0223}
\definecolor{litclr}{HTML}{1E0AFF}
\definecolor{gclr}{HTML}{000000}
\definecolor{lgclr}{HTML}{23FF00}
\tikzset{
lge/.style={circle, draw=black!80, fill=litgeclr!40, very thick, minimum size=8mm, outer sep=0},
ulge/.style={circle, draw=black!80, fill=unlitgeclr!40, very thick, minimum size=8mm,outer sep=0},
l/.style={circle, draw=black!80, fill=litclr!40, very thick, minimum size=1mm,outer sep=0},
ul/.style={circle, draw=black!80, fill=unlitclr!40, very thick, minimum size=1mm,outer sep=0},
g/.style={circle, draw=black!80, fill=gclr!40, very thick, minimum size=1mm,outer sep=0},
lg/.style={circle, draw=black!80, fill=lgclr!40, very thick, minimum size=1mm,outer sep=0},
active/.style={circle, draw=black!80, fill=lgclr!40, very thick, minimum size=1mm,outer sep=0},
result/.style={circle, draw=black!80, fill=unlitgeclr!40, very thick, minimum size=1mm,outer sep=0},
gvnode/.style={inner sep=0pt, outer sep=5pt}
}

\begin{document}

\title{PauLie: Fast Classification of Pauli Dynamical Lie Algebras}

\author{
\IEEEauthorblockN{Oxana Shaya\,\orcidlink{0009-0000-4200-2037}\IEEEauthorrefmark{1},
Konstantin Golovkin\,\orcidlink{0009-0002-5045-4775}\IEEEauthorrefmark{2},
Mainak Roy\,\orcidlink{0009-0003-8314-9686}\IEEEauthorrefmark{3},
Vincent Russo\,\orcidlink{0000-0002-8952-8981}\IEEEauthorrefmark{4}}
\IEEEauthorblockA{\IEEEauthorrefmark{1}Institute for Information Processing (tnt/L3S), Leibniz Universit\"at Hannover, Germany\\
Email: \href{mailto:shayaoxana@gmail.com}{shayaoxana@gmail.com}}
\IEEEauthorblockA{\IEEEauthorrefmark{2}Independent Researcher, Russia}
\IEEEauthorblockA{\IEEEauthorrefmark{3}Tata Institute of Fundamental Research, Mumbai, India}
\IEEEauthorblockA{\IEEEauthorrefmark{4}Unitary Foundation, USA}
}
\maketitle

\begin{abstract}
The dynamical Lie algebra (DLA) governs the controllability,
expressibility, and simulation complexity of a quantum system.
Explicitly computing it has been a major
computational bottleneck: brute-force Lie closure scales exponentially in the number of qubits $n$.
Many applications, however, consult only the isomorphism type of the DLA.
We introduce PauLie, an open-source framework that decides this
isomorphism type for DLAs generated by arbitrary Pauli strings, building
on the anticommutation-graph reduction of Aguilar et al.\ PauLie runs in $O(n|\G|\max(n,|\G|))$
time, where $|\G|$ is the number of generators, turning DLA
classification into a routine preprocessing step.
We demonstrate its use as a structural oracle for routing Lie-algebraic
simulation and Cartan decomposition, diagnosing barren plateaus in
variational quantum algorithms, and engineering universal Pauli string
generator sets with optimal generation rate.
\end{abstract}


\section{Introduction}
As quantum devices scale, efficient tools for compilation and controllability analysis become increasingly important.
The continuous-time evolution of a quantum system traces a trajectory in Hilbert space whose reachable set is determined by the underlying control Hamiltonian \cite{Zeier_2011,DAlessandro2007, Jurdjevic1972}.
For finite-dimensional closed quantum systems with piecewise-continuous controls, the set of reachable unitary operations is associated with a connected Lie group generated by the control Hamiltonian.
The tangent space to this group, the real span of the Hamiltonian terms and their nested commutators, is the dynamical Lie algebra (DLA).
The DLA's structure dictates a system's symmetries, expressibility, and controllability.
Explicitly determining a system's DLA has been a severe computational
bottleneck.
Brute-force methods build a basis by closing the generators under nested
commutators, and the dimension of that basis is generically exponential
in $n$: for a universal generator set it is $4^n-1$.
Any method that returns a basis therefore costs exponential time and
memory on such systems, however the closure is organized.
Recently there have been several breakthroughs. The DLAs of 2-local spin systems have been classified \cite{Wiersema_2024, kökcü2024classificationdynamicalliealgebras}. This line of research culminated in a classification
of DLAs generated by arbitrary sets of Pauli strings \cite{aguilar2024classificationpauliliealgebras}. 

PauLie \cite{PauLieRepo} turns this classification into a
general-purpose software framework built around an efficient
graph-theoretic reduction algorithm.
The algorithm achieves $O(n|\G|\max(n,|\G|))$ scaling with respect to the
number of qubits $n$ and number of generators $|\G|$, and improves upon
the concurrently proposed algorithm \cite{cuypers_2026} whenever
$|\G| \neq n$.
PauLie thus makes DLA analysis routinely available to quantum information scientists and engineers, who can use it in several ways:
First, as a key ingredient for Hamiltonian simulation 
and circuit synthesis \cite{wierichs2025recursivecartandecompositionsunitary}. 
Second, to identify the scope of efficient classical algorithms to simulate quantum systems with a small DLA \cite{Goh_2025, cerezo2023does}.
Third, to infer expressibility and barren-plateau behavior in algorithms like QAOA \cite{kazi2024analyzing, Ragone:2023qbn, Fontana:2023mgx}.
The framework can be used to study how the DLA changes by changing the generator set \cite{allcock2025generatingdirectpowersdynamical} and therefore has relevance in quantum system engineering. We demonstrate this by concretely identifying universal generator sets with optimal generation rate.  
More generally, graph-based representations of commutation relations between quantum observables appear in several areas of quantum information theory. For example, the anticommutation graph formalism provides a graph-theoretic description of qudit observable families \cite{makuta2025frustrationgraphformalismqudit}.
Moreover, we extended our framework to capture properties of the commutator graph which are related to quantities in quantum chaos and complexity~\cite{west2025graphtheoreticapproachchaoscomplexity}.

\section{Technical Background}\label{sec:prelims}

To understand PauLie's classification algorithm, we formalize the relationship between quantum dynamics, Lie algebras, and graph-theoretic representations.
\subsection{Quantum Dynamics and Matrix Lie Algebras}
A state of an isolated physical system is represented at a fixed time $t$ by a vector $\ket{\psi}$ in a Hilbert space $\mathcal{H}$, that is, a complex vector space endowed with an inner product.
The time evolution of the state is determined by the Schrödinger equation $i\frac{d}{dt} \ket{\psi}  = H \ket{\psi}$, where we set $\hbar = 1$.
$H$ is the Hamiltonian that describes the observable corresponding to the total energy of the system.
An observable is a Hermitian linear map acting on $\mathcal{H}$.
The solution to this differential equation with initial state $\ket{\psi_0}$ and time-independent Hamiltonian $H$ is $\ket{\psi(t)} = e^{-iHt} \ket{\psi_0}$.
For a Hamiltonian that decomposes as a sum $H = \sum_j h_j$, the terms $h_j$ typically do not commute, so $e^{-iHt}$ does not factor directly into a product of single-term exponentials. The Lie-Trotter formula resolves this by approximating
\begin{equation*}
  e^{-iHt} = \lim_{m \to \infty} \Big( \prod_j e^{-i h_j t/m} \Big)^{m}.
\end{equation*}
We specify a set of Hermitian operators $\mathcal{G} = \{h_j\}_{j\in J}$ that describe the ways in which we can control the quantum system. Depending on the context, the $h_j$ may be interaction terms of a Hamiltonian one wishes to simulate~\cite{Lloyd1996,SommaOrtizGubernatisKnillLaflamme2002}, control Hamiltonians of a quantum device~\cite{DAlessandro2021,AlbertiniDAlessandro2002}, or generators of a variational ansatz~\cite{WeckerHastingsTroyer2015}.
The reachable states are then of the form $\ket{\psi(\vec\theta)} = e^{-ih_{l_m} \theta_m} \dots e^{-ih_{l_1} \theta_1} \ket{\psi_0}$ for some index sequence $(l_1,\dots,l_m)\in J^m$ and parameters $\vec\theta \in \mathbb{R}^m$, which may be interpreted as evolution times, control amplitudes, or variational angles.
The matrix Lie algebra $\mathfrak{g}$ associated with $\mathcal{G}$, that is, the smallest real subspace of anti-Hermitian matrices containing $\{ih_j\}_{j\in J}$ and closed under the matrix commutator $\comm{h}{g} = hg-gh$, is called the dynamical Lie algebra (DLA)~\cite{Larocca2022DLA,Cerezo2021VQA}.
%The application of non-linear control theory to continuous group manifolds, laying the mathematical groundwork for assessing controllability via Lie algebras, was first introduced through the parallel foundational works of Brockett \cite{Brockett1972} and Jurdjevic and Sussmann \cite{Jurdjevic1972}. These Lie-algebraic principles were subsequently formalized specifically for finite-level quantum systems by D'Alessandro as the field of quantum control \cite{DAlessandro2007}.
By the Baker-Campbell-Hausdorff formula,
\begin{align*}
e^A e^B
  &= \exp\!\Big(
       A + B + \tfrac12\comm{A}{B}
       + \tfrac1{12}\comm{A}{\comm{A}{B}} \\
  &\qquad
       - \tfrac1{12}\comm{B}{\comm{A}{B}}
       + \cdots
     \Big),
\end{align*}
the elements in the Lie algebra generate precisely the unitaries that are reachable~\cite{JurdjevicSussmann1972}.
The classical compact simple Lie algebras are:
\begin{align*}
    \mathfrak{su}(d) &= \{x \in \mathbb{C}^{d\times d} \mid x = -x^\dagger, \tr(x) = 0\}\\
    \mathfrak{so}(d) &= \{x \in \mathbb{R}^{d\times d} \mid x = -x^T  \}\\
    \mathfrak{sp}(d) &= \{ x \in \mathfrak{su}(2d) \mid x =- \Omega x^T \Omega^T \}
\end{align*}
where the symplectic form is denoted as $\Omega = \begin{pmatrix} 0 & \mathbb{I}_d \\ - \mathbb{I}_d & 0 \end{pmatrix}$.
For a system of $n$ qubits, the Hilbert space is $\mathcal{H} = \mathbb{C}^{2^n}$.
A generator set whose DLA equals $\mathfrak{su}(2^n)$ is called universal, meaning the system is fully controllable.

\subsection{Graph-Theoretic Mapping of Lie Algebras}\label{subsec:graph_basics}
PauLie avoids dense matrices by mapping the generation of Pauli Lie algebras to a graph reduction problem \cite{aguilar2024classificationpauliliealgebras}.
The algebraic structure is extracted by analyzing the anticommutation relations of the generators.

\begin{definition}[Anticommutation Graph]
  The anticommutation graph $\Gamma = (\mathcal{G}, E)$ has the generating set $\mathcal{G}$ as its vertex set.
Undirected edges $E$ connect vertices whose corresponding Pauli operators anti-commute: $E = \{ (P, Q) \mid \{P, Q\} = 0 \}$.
\end{definition}

Any connected anticommutation graph can be reduced to a canonical representative by operations that alter the graph topology but leave the resulting DLA invariant.

\begin{definition}[Lightning]\label{def:lightning}
    The lightning of an anticommutation graph with respect to a vertex $V$ 
    is the induced subgraph of $\mathcal{G}\setminus \{V\}$ with binary labeling of each vertex $W\neq V$ as either \emph{lit} if $\{V,W\}=0$ or else \emph{unlit}.
\end{definition}

\begin{figure}[htbp]
    \centering
    \resizebox{\columnwidth}{!}{%
\begin{tikzpicture}
    \begin{scope}[local bounding box=s1]
        \node[lg](e) at (0,0){};
        \node[g,above right=1cm and 0.4cm of e](a){};
        \node[g,below right=1cm and 0.4cm of e](b){};
        \node[g,right=1.5cm of e](c){};
        \node[g,above right=0.5cm and 0.2cm of c](d){};
        \node[g,below right=0.5cm and 0.2cm of c](f){};
        \node[g,above right=0.4cm and 1.2cm of c](g){};
        \node[g,below right=0.4cm and 1.2cm of c](h){};
        \graph{(e) -- {(a),(b),(c),(d),(f)};(c) --{(a),(b),(d),(f),(g),(h)};(d)--{(a),(f),(g),(h)};(f)--{(b),(g),(h)};};
        \draw[->,red,rotate=-30] (d.north) arc [start angle=40,end angle=231,x radius=1.2cm,y radius=1.3cm];
        \node[draw=none,fill=none,fit=(a) (b) (c) (d) (e) (f) (g) (h)]{};
        \node[above left of=e,node distance=5mm] {$V$};
        \node[above right of=d,node distance=5mm] {$W$};
    \end{scope}
    \begin{scope}[local bounding box=s2,xshift=6cm, yshift=0.7cm]
        \node[l](a){};
        \node[l,below=1cm of a](b){};
        \node[l,below right=0.25cm and 0.5cm of a](c){};
        \node[l,above right=0.5cm and 0.2cm of c](d){};
        \node[l,below right=0.5cm and 0.2cm of c](f){};
        \node[ul,above right=0.4cm and 1.2cm of c](g){};
        \node[ul,below right=0.4cm and 1.2cm of c](h){};
        \graph{(c) --{(a),(b),(d),(f),(g),(h)};(d)--{(a),(f),(g),(h)};(f)--{(b),(g),(h)};};
        \begin{pgfonlayer}{background}
            \node[fit=(a) (b) (c) (d) (f) (g) (h)]{};
        \end{pgfonlayer}
        \node[above right of=d,node distance=5mm] {$W$};
    \end{scope}
    \begin{scope}[local bounding box=s3,yshift=-5cm]
        \node[lg](e) at (0,0){};
        \node[g,above right=1cm and 0.4cm of e](a){};
        \node[g,below right=1cm and 0.4cm of e](b){};
        \node[g,right=1.5cm of e](c){};
        \node[g,above right=0.5cm and 0.2cm of c](d){};
        \node[g,below right=0.5cm and 0.2cm of c](f){};
        \node[g,above right=0.4cm and 1.2cm of c](g){};
        \node[g,below right=0.4cm and 1.2cm of c](h){};
        \graph{(e) -- {(b),(d),(g),(h)};(c) --{(a),(b),(d),(f),(g),(h)};(d)--{(a),(f),(g),(h)};(f)--{(b),(g),(h)};};
        \node[draw=none,fill=none,fit=(a) (b) (c) (d) (e) (f) (g) (h)]{};
        \node[above left of=e,node distance=5mm] {$V$};
        \node[above right of=d,node distance=5mm] {$W$};
    \end{scope}
    \begin{scope}[local bounding box=s4,yshift=-4.3cm,xshift=6cm]
        \node[ul](a){};
        \node[l,below=1cm of a](b){};
        \node[ul,below right=0.25cm and 0.5cm of a](c){};
        \node[l,above right=0.5cm and 0.2cm of c](d){};
        \node[ul,below right=0.5cm and 0.2cm of c](f){};
        \node[l,above right=0.4cm and 1.2cm of c](g){};
        \node[l,below right=0.4cm and 1.2cm of c](h){};
        \graph{(c) --{(a),(b),(d),(f),(g),(h)};(d)--{(a),(f),(g),(h)};(f)--{(b),(g),(h)};};
        \begin{pgfonlayer}{background}
            \node[fit=(a) (b) (c) (d) (f) (g) (h)]{};
        \end{pgfonlayer}
        \node[above right of=d,node distance=5mm] {$W$};
    \end{scope}
    \draw[-to, thick] (4.5,-2) -- (4.5,-3) node [right=1mm,midway,text width=3cm]{Contract$(V,W)$};
    \draw[-to, thick] ([xshift=5mm]s1.east) -- ([xshift=-5mm]s1.east -| s2.west) node [above=1mm,midway,text width=3cm,text centered]{$\Gamma(V)$};
    \draw[-to, thick] ([xshift=5mm]s3.east) -- ([xshift=-5mm]s3.east -| s4.west) node [above=1mm,midway,text width=3cm,text centered]{$\Gamma(V)$};
\end{tikzpicture}
}
    \caption{
    The anticommutation graph of $
\mathcal{G}= \{\,YXIZ,\; ZZII,\; XIII,\; ZIII,\; YIIX,\; YIIZ,\; YYIX,\; YYXX\,\}
$ is drawn in the upper left. Its lightning $\Gamma(V)$ is drawn in the upper right where the blue marks lit vertices. 
    The result of contracting  $V=YIIX $ by $W = ZIII$ to $XIIX$ is drawn in the lower left. The lightning after the contraction is shown in the lower right.
    }
    \label{fig:contraction}
\end{figure}
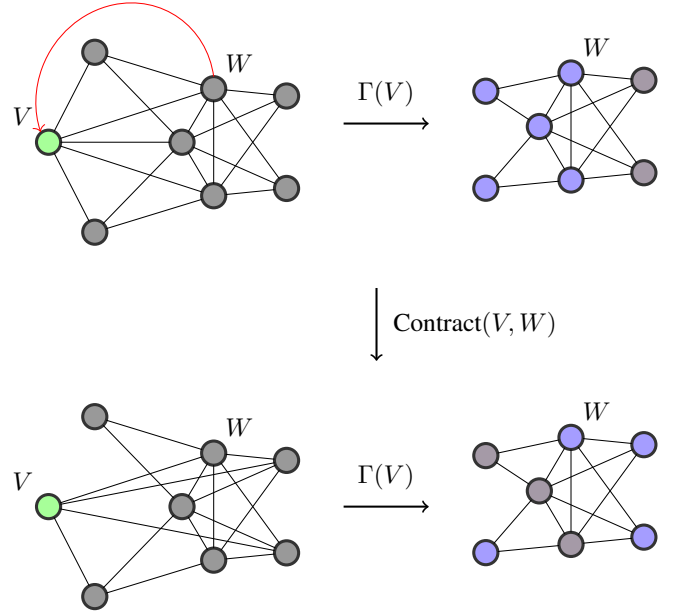

\begin{definition}[Contraction]\label{def:contraction}
    A contraction along an edge $(V, W)$ replaces the generator $V$ with the scaled commutator $X \propto \comm{V}{W}$.
\end{definition}
Structurally, $X$ remains adjacent to $W$, while every other vertex toggles
its adjacency to the contracted vertex precisely if it is lit in the
lightning with respect to $W$: common neighbours of $V$ and $W$ lose their
edge, and vertices adjacent to $W$ alone gain one.
We illustrate a contraction in Figure~\ref{fig:contraction}.

By systematically applying contractions the graph is mapped to a canonical structure: a line graph (Type~A) or a \emph{starlike} graph (Type~B).
For Type~A graphs, \(n_L\) denotes the number of vertices in the line and \(n_c\) the number of single vertices attached to the penultimate vertex. 
A starlike graph consists of a central vertex (the core) connected to a set of disjoint simple path graphs (legs).  If the graph has no legs longer than \(2\), it is of Type~B1. If it has exactly one leg of length \(4\) and none of length \(3\), it is Type~B2; otherwise it is Type~B3. For starlike graphs  \(n_c+1\) denotes the number of legs of length \(1\), and \(n_2\) the number of legs of length \(2\).

\begin{figure}[htbp]
    \centering
    \resizebox{\columnwidth}{!}{%
\begin{tikzpicture}[inner sep=0pt]
    \node[lge] (O) {$O$};
    \node[ulge] (L11) [right =of O] {$L^{(1)}_4$};
    \node[ulge] (L12) [right =of L11]{$L^{(2)}_4$};
    \node[ulge] (L13) [right =of L12]{$L^{(3)}_4$};
    \node[ulge] (L21) [above left =of O]{$L^{(1)}_1$};
    \node[ulge] (L31) [below left =of O]{$L^{(1)}_2$};
    \node[ulge] (L41) [left =of O]{$L^{(1)}_3$};
    \node[ulge] (L42) [left =of L41]{$L^{(2)}_3$};

    \draw[-] (O) to (L11) to (L12) to (L13);
    \draw[-] (O) to (L21);
    \draw[-] (O) to (L31);
    \draw[-] (O) to (L41) to (L42);
\begin{scope}[on background layer]
    \node [inner sep=1mm,fill=yellow!40,draw=black!40,rounded corners,fit=(L11) (L12) (L13)] {};
    \node [inner sep=1mm,fill=yellow!40,draw=black!40,rounded corners,fit=(L21)] {};
    \node [inner sep=1mm,fill=yellow!40,draw=black!40,rounded corners,fit=(L31)] {};
    \node [inner sep=1mm,fill=yellow!40,draw=black!40,rounded corners,fit=(L41) (L42)] {};
\end{scope}
\end{tikzpicture}
}
    \caption{An example of a canonical starlike graph of type B3. The central vertex is $O$ and leg nodes are denoted by $L_i^{(j)}$ where $i$ counts the leg and $j$ counts the node inside the leg. Here $n_c =1$, $n_2=1$ and the associated dynamical Lie algebra is $\mathfrak{su}(8) \oplus \mathfrak{su}(8)  $.}
    \label{fig:starlike}
\end{figure}
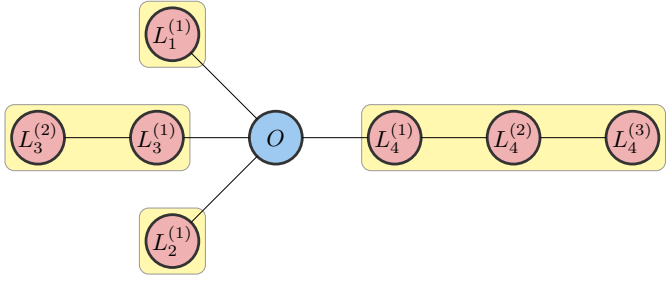

An example of such a canonical graph is presented in Figure~\ref{fig:starlike}.

\section{Pauli DLA Classification Algorithm}\label{sec:classification}

PauLie implements the classification of Aguilar et
al.~\cite{aguilar2024classificationpauliliealgebras} as an optimized graph
algorithm.
The algorithm operates directly on the anticommutation graph of the
generator set.
Unlike methods that rely on invariants of an associated quadratic
space \cite{cuypers_2026}, it achieves $O(n|\G|\max(n,|\G|))$ scaling
through graph transformations.
The algorithm assumes no locality; its input is an arbitrary set of Pauli strings.

By \cite[Section VI]{aguilar2024classificationpauliliealgebras}, the connected components of the anticommutation graph correspond to the direct sum of the Lie algebras generated by each connected component. Therefore, PauLie first isolates these components and independently canonicalizes each component. The algorithm iteratively builds the canonicalized graph by adding vertices in a loop and utilizing contractions to ensure the canonical property is maintained in every iteration. In the following description, we count the number of \emph{basic operations}, by which we mean multiplication and commutation of two Pauli strings.

\paragraph{Canonicalization Routine}
Vertices are processed from a stack, initialized with a DFS preorder of
the component. The routine maintains a single invariant: after every
iteration, the vertices processed so far form a canonical graph---a central
vertex with legs sorted by length---that generates the same DLA as the
generators it was built from. We describe an iteration from the perspective
of the vertex $V$ being added; lit and unlit refer to the lightning with
respect to $V$ (Definition~\ref{def:lightning}), and we call the longest
leg the \emph{long leg}. Contractions (Definition~\ref{def:contraction})
are the basic move; occasionally the routine instead multiplies a vertex
already in the graph by other generators, which likewise preserves the DLA.
Each iteration reduces the lightning until $V$ is adjacent only to vertices
at which the canonical form admits an attachment, and attaches it there:
\begin{enumerate}
    \item \textbf{Building the core:} The first vertex becomes the central
    vertex. While the graph has fewer than two legs, each added vertex is
    attached directly to the centre: if it is lit on the existing leg,
    contractions with that leg and with the centre first remove the unwanted
    adjacency, and $V$ then forms a new leg of length 1.
    \item \textbf{Handling legs of length 1:} All legs of length 1 must be
    in the same state, either all lit or all unlit, so that one of them can act
    as a \emph{representative} for the rest. If instead some leg $P$ is lit
    and some leg $Q$ is unlit, we apply the reduction behind the first part
    of \cite[Theorem 4]{aguilar2024classificationpauliliealgebras}:
    multiplying every other lit vertex by the product $PQ$ preserves the
    generated algebra and turns its light off, so $V$ ends up adjacent to
    $P$ alone, attaches there, and the iteration ends with a new leg of
    length 2. If this gives a type-A graph a second leg of length at least
    2, the graph becomes type B; since type B admits no leg longer than 4,
    the long leg is trimmed to length 4, and the removed vertices are pushed
    back onto the stack to be reprocessed later (a \emph{re-add}). When the
    length-1 legs already agree, contractions make $V$ unlit on the
    representative---hence on all of them---and lit on the central vertex.
    If no vertex of any longer leg is lit either, $V$ simply becomes a new
    leg of length 1 and the iteration ends.
    \item \textbf{Transferring the lightning to the long leg:} Next, the
    remaining light is concentrated on the long leg. Each length-2 leg with
    a lit vertex is cleared by contractions with its own vertices, the first
    vertices of the long leg, and the representative; in exchange, the light
    moves onto the long leg. If after this only the central vertex is lit,
    $V$ attaches to it as a new leg of length 1 and the iteration ends.
    \item \textbf{Handling the long leg:} Only the central vertex and the
    long leg now carry light. If the central vertex is unlit, contracting
    $V$ successively with the long-leg vertices, from the first lit position
    down to the first vertex, lights it. (One shortcut comes first: in a
    type-A graph whose long leg has length at least 5, with only the last
    vertex lit, $V$ is appended at the end of the leg directly.) While at
    least two long-leg vertices remain lit, an inner loop contracts $V$ with
    contiguous runs of the leg; every pass moves the lit positions towards
    the end of the leg, so eventually exactly one vertex is lit. Precomputed
    prefix products make each of these contractions cost $O(1)$ basic
    operations, as exploited in the runtime analysis below. Two cases
    remain.
    \begin{enumerate}
        \item \textbf{First vertex is lit:} Contracting along the whole leg
        moves the light to its end, and $V$ is appended there. The exception
        is type B2, whose long leg must not grow: there, the move of
        \cite[Lemma 3]{aguilar2024classificationpauliliealgebras} attaches
        $V$ to the central vertex as a new leg of length 1 instead.
        \item \textbf{Some other vertex is lit:} The long leg is broken just
        before the lit position: the preceding vertex absorbs $V$ and the
        representative, and the tail of the leg, now headed by $V$, becomes
        a second leg attached to the centre. If the graph was type A and is
        left with two legs of length at least 2, it becomes type B, and the
        surplus is trimmed---the longer leg to length 4, the other to length
        2---with all removed vertices re-added via the stack. A re-add is
        not an actual removal; it is a device to generate, over the next
        iterations, the contraction sequence that returns the graph to
        canonical form.
    \end{enumerate}
\end{enumerate}

At the end of the procedure, we must remove all dependent Pauli strings for the correct classification. By \cite[Section IV.A]{aguilar2024classificationpauliliealgebras}, we only need to perform Gaussian elimination over the legs of length 1.

\begin{algorithm}[htbp]
\caption{PauLie Canonicalizer}
\label{algo:main}
\label{algo:canonicalizer}
\begin{algorithmic}[1]
\Require Pauli generator set $\mathcal{G}$
\State $S \gets$ Vertex stack
\State $C \gets \textbf{null}$ \Comment{Central vertex}
\State $L \gets []$ \Comment{List of graph legs}
\While{$S$ is not empty}
    \State $V \gets \text{pop}(S)$
    \If{$C$ is \textbf{null}}
        \State $C \gets V$
        \State \Continue
    \ElsIf{$\text{length}(L) < 2$}
        \State Build core by contracting $V$ into $C$ and $L_1$
        \State \Continue
    \ElsIf{length-1 legs in $L$ have differing lit states}
        \State Use \cite[Theorem 4]{aguilar2024classificationpauliliealgebras} to create a length-2 leg
        \While{length of longest leg in $L > 4$}
            \State Pop last vertex from longest leg in $L$
            \State Push vertex to $S$
        \EndWhile
        \State \Continue
    \EndIf
    \State $\omega \gets$ any length-1 leg
    \State Make $\omega$ unlit and $C$ lit
    \If{only $C$ is lit}
        \State Add $V$ to $L$ as a length-1 leg
        \State \Continue
    \EndIf
    \If{any length-2 leg in $L$ has a lit vertex}
        \State Transfer lightning away from length-2 legs
    \EndIf
    \State Reduce lightning on the longest leg
\EndWhile
\State Perform Gaussian elimination over length-1 legs in $L$

\State \Return Canonical graph constructed from $C$ and $L$
\end{algorithmic}
\end{algorithm}

The specific geometric configuration of the resulting canonical graph uniquely identifies the DLA as one of four families, as detailed in Table~\ref{tab:canonical_types}. There is one exception, which is a single Pauli string. This case generates the algebra $\mathfrak{u}(1)$.

\begin{table}[htbp]
\centering
\renewcommand{\arraystretch}{1.6}
\setlength{\tabcolsep}{10pt}
\begin{tabular}{@{}ll@{}}
\toprule
\textbf{Type} & \textbf{Algebraic Isomorphism} \\ \midrule
 A & $\bigoplus_{i=1}^{2^{n_c}} \So(n_L + 1)$ \\ 
 B1 & $\bigoplus_{i=1}^{2^{n_c}} \Sp(2^{n_2})$ \\ 
 B2 & $\bigoplus_{i=1}^{2^{n_c}} \So(2^{n_2+3})$ \\ 
 B3 & $\bigoplus_{i=1}^{2^{n_c}} \Su(2^{n_2+2})$ \\ \bottomrule
\end{tabular}
\caption{Correspondence between the canonical graph type and the associated DLA. Type~A corresponds to line graphs, while Type~B corresponds to starlike graphs. }
\label{tab:canonical_types}
\end{table}

\paragraph{Runtime Analysis}

\emph{Comparison to prior work.} Standard numerical Lie-closure
algorithms perform rank checks on a $2^{2n}\times N$ basis matrix,
where $N$ is the current closure dimension;
even with recent optimizations~\cite{iiyama2025fast} they remain exponential in the worst case. 
The Lie closure is sufficient but not necessary to determine the isomorphism type. The concurrent algebraic
approach of~\cite{cuypers_2026} does decide the isomorphism type 
through $\mathbb{F}_2$ quadratic-space invariants in
$\mathcal{O}(\max(n,|\G|)^3)$ time. PauLie improves on that bound whenever $|\G|\neq n$ and
matches it at $|\G|=n$, as we show below.

\emph{Cost model.} Under PauLie's bit-packed symplectic encoding, a
Pauli string on $n$ qubits occupies $O(n/W)$ words, where $W$ is the
word size (typically $64$). Multiplication and commutation of
two Pauli strings reduce to bitwise operations on these arrays and
cost $O(n/W)$ each.

\emph{Cost per iteration.} The \emph{main loop} of Algorithm~\ref{algo:canonicalizer} is its outer \textbf{while} loop, and one execution of its body---popping a single vertex from the stack and processing it through Steps 1--4 of the canonicalization routine---is an \emph{iteration}. The cost of a single iteration breaks down as:
\begin{center}

\begin{tabular}{@{}lcccc@{}}
\toprule
\textbf{Step}                    & 1      & 2         & 3         & 4         \\ \midrule
\textbf{Cost (basic operations)} & $O(1)$ & $O(|\G|)$ & $O(|\G|)$ & $O(|\G|)$ \\
\bottomrule
\end{tabular}
\end{center}

The optimization in Step 4 relies on three facts: contractions reduce
to multiplications when the result is known to be non-zero; Pauli
strings are involutory; and coefficients are immaterial for the
classification. Together these let prefix products along the long leg
be precomputed once per iteration in $O(|\G|)$ time. Since a lit-vertex
reduction corresponds to contraction with a contiguous range of vertices,
the earlier facts mean we can compute this by just taking the product of
two precomputed prefix products. Thus each of the $O(|\G|)$ lit-vertex
reductions costs $O(1)$. Per-iteration total: $O(|\G|)$ basic operations.

\emph{Main-loop total.} The number of iterations equals the number of
vertices ever pushed onto the stack. The stack initially holds $|\G|$
vertices, and the algorithm can also push already-processed vertices back
onto it (\emph{re-adds}). These arise in two ways.

First, at the transition from type~A to type~B, which may occur at either
of two sites in the canonicalization routine (Step~2 and Step~4b). Once
the graph is type~B it cannot become type~A again, since there is no way
to remove length-2 legs, so the two sites share a single budget of at
most $|\G|$ re-adds, no vertex being re-added more than once.

Second, when the long leg of a type~B graph shrinks, from length $4$ to
$3$ or from $3$ to $2$. Here the re-added vertex is pushed onto the top
of the stack and popped in the following iteration, where it is adjacent
both to the end of the long leg and to the vertex attached to the core,
and is therefore attached as the second vertex of a length-2 leg in
$O(1)$ time. At most a constant number of such re-adds is produced per
iteration, so each can be charged to the iteration that produced it.

The main loop therefore runs $O(|\G|)$ iterations of cost $O(|\G|)$
together with $O(|\G|)$ iterations of cost $O(1)$, giving $O(|\G|^{2})$
basic operations.

\emph{Final dependency removal.} After canonicalization, we drop
length-1 leg generators that are products of others. The bit-packed
symplectic encoding gives a bijection between Pauli strings on $n$
qubits and $\mathbb{F}_2$ vectors of length $2n$, so dependency
removal is Gaussian elimination in $\mathbb{F}_2^{2n}$. We maintain the basis in reduced row-echelon form, indexed by pivot column so that the unique basis vector with a given pivot can be retrieved in $O(1)$. Reducing a
new vector requires repeated XORs against the basis vector sharing
its current pivot; since the pivot strictly advances with each XOR
and lies in $\{1,\ldots,2n\}$, reduction terminates in $O(n)$ XORs
per vector. With $O(|\G|)$ length-1 legs to process, the dependency-removal step
costs $O(n|\G|)$ basic operations.

\emph{Total.} Summing the two phases:
\begin{center}
\resizebox{\columnwidth}{!}{%
\begin{tabular}{@{}lll@{}}
\toprule
\textbf{Phase} & \textbf{Basic operations} & \textbf{Word operations} \\ \midrule
Main loop                & $O(|\G|^2)$           & $O(n|\G|^2)$           \\
Final dependency removal & $O(n|\G|)$            & $O(n^2|\G|)$           \\ \midrule
\textbf{Total}           & $O(|\G|\max(n,|\G|))$ & $O(n|\G|\max(n,|\G|))$ \\
\bottomrule
\end{tabular}
}
\end{center}
The word-operation column applies the cost model above: each basic operation (Pauli multiplication, commutation, or $\mathbb{F}_2^{2n}$ XOR) costs $O(n/W)$ word operations on the bit-packed symplectic encoding.  In total,
\[
T(n,|\G|) \;=\; \mathcal{O}\!\left({n|\G|(n+|\G|)}\right) = O(n|\G|\max(n,|\G|)) .
\]

\emph{Connectivity.} We may choose to initialize the stack with an arbitrary ordering of vertices, but this may lead to a situation where the vertex to be added is disconnected from the intermediate graph. In this situation, we use a queue instead of a stack, and thus we push the current vertex to the end of the queue. We do the same for re-adds. However, given that the cost of adding a new vertex is linear in the number of already added vertices, it is simple to show that no matter how many passes are required, since at least one vertex is added in each pass, and since the cost of skipping a vertex is $O(1)$, the total cost remains bounded by $O(n|\G|^2)$. In the worst case, a re-add requires us to rebuild almost the whole graph from scratch, but this only happens once, so the total cost continues to hold. Finally, it is possible to prove with some effort that initializing the stack with a DFS preorder ensures that connectivity is never lost. However, in light of the above fact, we have omitted this proof.

\paragraph{Example execution of algorithm}
We classify the generator set of Figure~\ref{fig:contraction} using the DFS preorder
$S = [ZZII,$ $YIIX,$ $YXIZ,$ $ZIII,$ $YYXX,$ $XIII,$ $YIIZ,$ $YYIX]$.
Table~\ref{tab:example_run} traces each iteration and Figure~\ref{fig:example} illustrates the corresponding graph transformations.
The final canonical graph has legs of lengths $\{1,1,2,3\}$, so by Table~\ref{tab:canonical_types} the algebra is $\Su(8)\oplus\Su(8)$ (type~B3 with $n_c=1$, $n_2=1$).

\begin{table*}[htbp!]
\centering
\caption{Trace of Algorithm~\ref{algo:canonicalizer} on the generators of Figure~\ref{fig:contraction}. Each block of rows corresponds to one popped vertex; sub-rows list contractions performed within that iteration. The notation $\cdot P$ denotes contraction with $P$, and $\cdot(P_1,\ldots,P_k)$ denotes a lightning sequence applied in order.}
\label{tab:example_run}
\setlength{\tabcolsep}{6pt}
\renewcommand{\arraystretch}{1.2}
\resizebox{\textwidth}{!}{%
\begin{tabular}{@{}l l l l@{}}
\toprule
Popped $V$ & Branch / sub-step & Operation & Effect \\ \midrule
$ZZII$ & Step 1: seed core & --- & $C \gets ZZII$ \\ \midrule
$YIIX$ & Step 1: $<\!2$ legs, attach to core & --- & first length-1 leg \\ \midrule
$YXIZ$ & Step 1: $<\!2$ legs, attach to core & $\cdot YIIX,\ \cdot ZZII$ & $YXIZ \to ZYIY$; second length-1 leg \\ \midrule
$ZIII$ & extend $YIIX$ leg & --- & longest leg length 2 (type A) \\ \midrule
$YYXX$ & extend longest leg & --- & longest leg length 3 (type A) \\ \midrule
$XIII$ & Step 2: skipped (legs share lit state); normalize & $\cdot ZZII$ & $XIII \to YZII$ \\
       & Step 3: skipped (no length-2 legs)              & ---            & --- \\
       & Step 4: long leg has lit vertices               & $\cdot YYXX$   & $YZII \to IXXX$ \\
       & Step 4b: lightning relocates $ZIII$             & $\cdot(YIIX, ZZII, IXXX, ZYIY, ZZII, YIIX)$ & $ZIII \to IZXZ$; type B, two length-2 legs \\ \midrule
$YIIZ$ & Step 3: length-2 leg has lit vertex             & $\cdot YIIX$   & $YIIZ \to IIIY$ \\
       & Step 3 (cont.)                                  & $\cdot IXXX$   & $IIIY \to IXXZ$, central lit \\
       & Step 3 (cont.)                                  & $\cdot(ZZII, IZXZ, YIIX, ZYIY, ZZII)$ & $IXXZ \to XIIZ$ \\
       & Step 4a: first vertex of long leg lit           & $\cdot IXXX,\ \cdot YYXX$ & $XIIZ \to ZZIZ$, attached to $YYXX$ \\ \midrule
$YYIX$ & Step 4: light central vertex                    & $\cdot IXXX$   & $YYIX \to YZXI$ \\
       & Step 4 (cont.): second-last of long leg lit     & $\cdot YYXX$   & $YZXI \to IXIX$ \\
       & Step 4b: lightning relocates $IXXX$             & $\cdot(ZZII, IXIX, ZYIY, ZZII)$ & $IXXX \to ZYXY$; legs $\{1,1,2,3\}$, type B3 \\
\bottomrule
\end{tabular}%
}
\end{table*}

\begin{figure*}[p]
    \centering
    \input{diagrams/example_1.tikz}
    \caption{A visual walkthrough of the example in Table~\ref{tab:example_run}.}
    \label{fig:example}
\end{figure*}

\section{Software Design and Architecture}\label{sec:software}
PauLie's architecture decouples the memory representation of operators from their algebraic properties, so that the user-facing API tracks the underlying mathematical formalism while the bit-level backend can be optimized or ported independently. Figure~\ref{fig:class_diagram} summarizes the principal classes; the description below walks through the two layers it depicts.

\begin{figure*}[htb!]
  \centering
  \tikzset{external/export=false}
  \includegraphics[width=\textwidth]{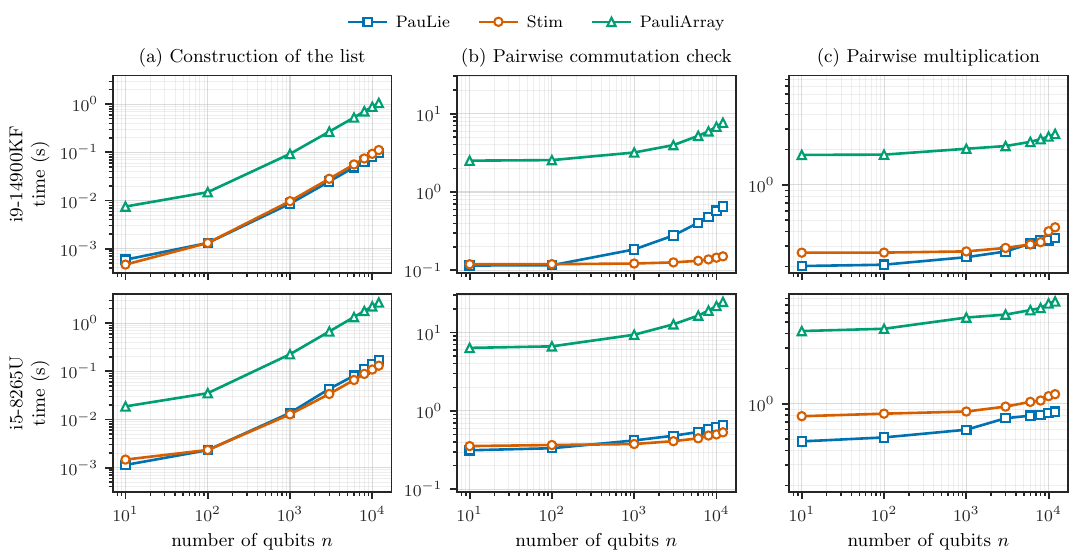}
  \caption{Backend primitives as a function of the number of qubits, on two
processors (rows), for a list of $1000$ random Pauli strings: construction of
the list (a), and all $\binom{1000}{2}$ pairwise commutation checks
(b) and pairwise multiplications (c). Each point is the minimum over $100$
repetitions on the i9-14900KF and over $30$ on the i5-8265U. Both axes are
logarithmic and the two rows share a vertical scale within each column.
PauLie leads on multiplication at all but two measured points and is level
with \texttt{Stim} on construction, the small residual lead flipping between
the two processors; on commutation it is overtaken by \texttt{Stim} for
$n \gtrsim 10^{3}$, where \texttt{Stim}'s cost on the i9-14900KF is nearly
flat in $n$.}
  \label{fig:backend}
\end{figure*}

The operator and collection layer lives in \texttt{paulie.common}. \texttt{PauliString} stores an $n$-qubit operator as a single bit array of length $2n$ in its symplectic representation, backed by the dedicated \texttt{pauliebits} library, so that multiplication, commutation, and the adjoint action all reduce to bitwise XOR and parity computations---producing the $O(n/W)$ basic-operation cost analyzed in Section~\ref{sec:classification}.
Figure~\ref{fig:backend} benchmarks these primitives against \texttt{Stim}~\cite{gidney2021stim} and \texttt{PauliArray}~\cite{dion2024efficiently}.

\texttt{PauliStringCollection} models a generator set and provides two entry points to the pipeline: \texttt{get\_algebra} returns the DLA as a string label, while \texttt{classify} returns the full \texttt{Classification} object for callers that need the canonical graphs or the DLA dimension. Construction across both classes is mediated by an overloaded \texttt{get\_pauli\_string} factory.

The classification pipeline lives in \texttt{paulie.classifier}. The \texttt{build\_canonical\_graph} method of the \texttt{Canonicalizer} class consumes a DFS-ordered vertex stack from one connected component and produces a \texttt{Morph}---a typed canonical-graph object that records the legs and exposes the type label and leg counts needed to read off the algebra family from Table~\ref{tab:canonical_types}. \texttt{Classification} aggregates the per-component \texttt{Morph}s, as the vertices of the canonical graph, into the direct-sum decomposition. Downstream functionality such as \texttt{get\_optimal\_universal\_generators} lives in a separate \texttt{application} subpackage that consumes only the public interfaces of \texttt{PauliStringCollection} and \texttt{Classification}, so new applications can be added without modifying core classes and the bit-packed backend can be swapped (for example for a Cython or Rust implementation) without touching them.

\begin{figure*}[t]
\centering
\tikzset{external/export=false}
\begin{tikzpicture}[
  every node/.style={font=\small},
  class/.style={
    rectangle split,
    rectangle split parts=2,
    rectangle split part align=left,
    rectangle split draw splits=true,
    draw, thick, rounded corners=1pt,
    text width=4.2cm, inner sep=4pt, align=left
  },
  composes/.style={{Diamond[open,length=3mm,width=2.5mm]}-, thick},
  uses/.style={-{Stealth[length=2.5mm]}, dashed, thick},
  produces/.style={-{Stealth[length=2.5mm]}, thick}
]

% Top: collection holds many strings
\node[class] (pcoll) {%
  \textbf{\texttt{PauliStringCollection}}
  \nodepart{second}
  {\footnotesize Generator set; entry points \texttt{classify()} and \texttt{get\_algebra()}.}};

\node[class, right=3.0cm of pcoll] (pstring) {%
  \textbf{\texttt{PauliString}}
  \nodepart{second}
  {\footnotesize Pauli operator with symplectic bit-packed representation.}};

% Bottom: classification pipeline
\node[class, below=1.6cm of pcoll] (canon) {%
  \textbf{\texttt{Canonicalizer}}
  \nodepart{second}
  {\footnotesize Implements Algorithm~\ref{algo:main} via graph contractions on one connected component.}};

\node[class, right=1.5cm of canon] (morph) {%
  \textbf{\texttt{Morph}}
  \nodepart{second}
  {\footnotesize Canonical graph of one connected component; carries the type label and leg structure.}};

\node[class, right=2.5cm of morph] (classif) {%
  \textbf{\texttt{Classification}}
  \nodepart{second}
  {\footnotesize DLA as direct sum $\bigoplus_i \mathfrak{g}_i$, assembled across components.}};

% Edges
\draw[composes] (pcoll.east) -- node[above, font=\scriptsize]{generators} (pstring.west);
\draw[uses] (pcoll.south) -- node[right=2pt, font=\scriptsize]{\texttt{classify()} runs the pipeline} (canon.north);
\draw[produces] (canon.east) -- node[above, font=\scriptsize]{produces} (morph.west);
\draw[composes] (classif.west) -- node[above, font=\scriptsize]{per subgraph} (morph.east);

\end{tikzpicture}
\caption{Class diagram of the PauLie classification pipeline. A \texttt{PauliStringCollection} holds a generator set of \texttt{PauliString}s. Calling \texttt{classify()} on the collection instantiates a \texttt{Canonicalizer} and a \texttt{Classification}; for each connected component of the anticommutation graph, the \texttt{Canonicalizer} (Algorithm~\ref{algo:main}) builds a \texttt{Morph} that is added to the \texttt{Classification}. The \texttt{Classification} is stored as an attribute of the \texttt{PauliStringCollection} and assembles the per-component \texttt{Morph}s, as the vertices of the canonical graph, into the direct-sum decomposition of the DLA. A diamond at one end of a connection marks the container side of a ``contains-many'' relationship (a \texttt{PauliStringCollection} contains many \texttt{PauliString}s; a \texttt{Classification} contains many \texttt{Morph}s). The solid arrow marks the \texttt{Canonicalizer}'s \texttt{Morph} output, and the dashed arrow marks the runtime call that drives the pipeline.}
\label{fig:class_diagram}
\end{figure*}

\section{Benchmarking}\label{sec:benchmarking}
\begin{figure*}[t]
  \centering
  \tikzset{external/export=false}
  \includegraphics[width=.8\textwidth]{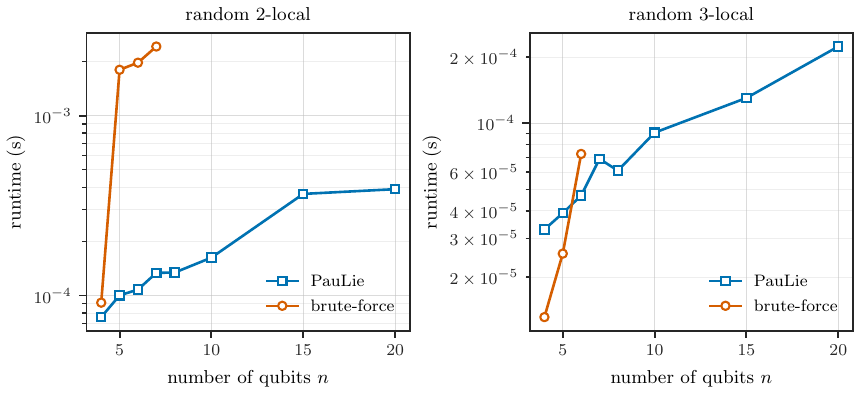}
  \caption{Runtime of PauLie's classification and of
brute-force closure on random $k$-local generators. The baseline returns
a basis and PauLie returns an isomorphism type. Each point is the minimum over $7$ repeated
timings. The
brute-force baseline is capped at 30 s per instance.}
  \label{fig:bench_random}
\end{figure*}

We validated the implementation by reproducing the tabulated results on DLAs generated by two-local Pauli strings \cite{Wiersema_2024} in \texttt{test\_classification.py}.
We then benchmarked PauLie against a brute-force Lie closure
algorithm on random $k$-local Pauli Hamiltonians with
$|\mathcal{G}| = (4-k)n$ terms.
The runtime comparison is shown in Figure~\ref{fig:bench_random}.
At $n = 20$ qubits PauLie completes the classification in
$390\,\mu\mathrm{s}$ for $k = 2$ and $223\,\mu\mathrm{s}$ for $k = 3$,
while the brute-force baseline already takes $2.4\,\mathrm{ms}$ at
$n = 7$.

\begin{figure*}
  \centering
  \includegraphics[width=0.6\linewidth]{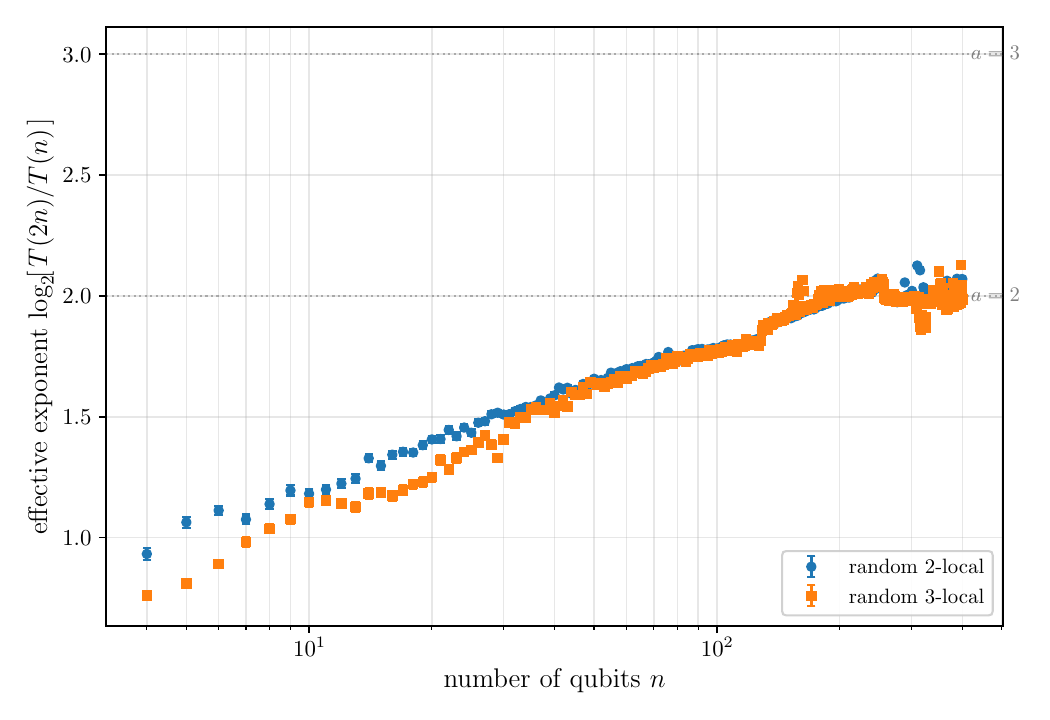}
  \caption{%
    \textbf{Empirical effective exponent of \texttt{get\_algebra} via the doubling-ratio method.}
    For each $n$ such that both $T(n)$ and $T(2n)$ have been measured, the quantity $\log_{2}\bigl[T(2n)/T(n)\bigr]$ converges to the exponent $a$ of any asymptotic scaling $T(n) \sim c\,n^{a}$.
    Runtimes are measured for random $k$-local Pauli Hamiltonians with $k \in \{2, 3\}$ and $n \in \{4, 5, \dots, 800\}$, using $100$ independent random instances per $n$; each instance is denoised by taking the minimum over $3$ repeated timings of \texttt{get\_algebra}.
    Error bars show propagated standard error of the mean of $T(n)$ and $T(2n)$.
    Dotted grey lines mark the values $a = 2$ and $a = 3$ for reference.
  }
  \label{fig:bench_scaling}
\end{figure*}

To characterise the empirical scaling of \texttt{get\_algebra} we employ the \emph{doubling-ratio} diagnostic, a non-parametric method for empirical complexity analysis \cite{sedgewick-algorithms, kapfhammer-expose}.
For any asymptotic scaling $T(n) \sim c\, n^{a}$, the ratio of runtimes at input sizes $n$ and $2n$ satisfies
\begin{equation}
  \log_{2}\!\left[\frac{T(2n)}{T(n)}\right] \xrightarrow{n \to \infty} a,
\end{equation}
so the effective exponent can be read directly off the $y$-axis as a function of $n$, without fitting a parametric model and without small-$n$ overhead affecting the large-$n$ estimate.
Figure~\ref{fig:bench_scaling} shows the measured ratios for random $k$-local Pauli Hamiltonians.
Across the measured range, the ratio rises monotonically and approaches~$2$ at the largest accessible $n$, remaining well below the $a = 3$ reference throughout.
This is consistent with the theoretical worst-case bound of $O(n|\G|\max(n,|\G|))$ and demonstrates that the typical-case runtime on random $k$-local Hamiltonians is substantially more favourable.

\section{Applications of DLA Classification}\label{sec:applications}

The classification is fast enough to be consulted inside larger workflows; this section collects such applications.
\subsection{Lie-Algebraic Simulation}
Synthesizing unitaries via exact Cartan decomposition requires a Lie algebra with a compact representation \cite{wierichs2025recursivecartandecompositionsunitary}.
Likewise, classical simulation of quantum systems is efficient in the polynomial-sized DLA regime \cite{Goh_2025}. 
Classifying the generator set first ensures that these algorithms are invoked only for polynomial-sized DLAs, while exponentially large algebras are handed to approximate simulation techniques.

\subsection{Generator-Set Engineering}
PauLie also supports the engineering of generator sets. 
Concretely, we identify generator sets consisting of Pauli strings with optimal generation rate for any dimension.
These are optimal in the sense that the number of Pauli strings generated in successive rounds of nested commutators grows maximally. As shown in \cite{Smith_2025}, optimal generating sets are characterized by having a fraction of anticommuting generator pairs close to $0.706$. Moreover, the minimal number of Pauli-string generators required to generate $\Su(2^n)$ is $2n + 1$. This allows the problem to be formulated in terms of the anticommutation graph. Optimal sets therefore correspond to graphs with approximately
\[
\left\lfloor 0.706 \cdot \binom{2n+1}{2} \right\rfloor
\]
edges.

The search starts from the minimal universal set of
\cite[Example~1]{Smith_2025}. The seed's anticommutation fraction lies far below the optimal $0.706$.
Repeatedly replacing a generator $P$ by its product $PQ$ with an
anticommuting neighbour $Q$ preserves minimality and universality while
rewiring the anticommutation graph, and replacements are applied until the
number of anticommuting pairs reaches the target above. The search is
implemented in \texttt{get\_optimal\_universal\_generators}.
Beyond such concrete questions on gate sets, PauLie can be employed to study theoretical algebraic questions. 
Allcock et al.
\cite{allcock2025generatingdirectpowersdynamical} demonstrated how to construct direct sums of multiple copies of a DLA with only logarithmically many extra qubits.
PauLie identifies the required anticommutation structures directly, so these direct-power constructions can be verified, and related open questions probed, numerically.

\subsection{Trainability of variational quantum algorithms}
In quantum machine learning, the DLA controls the onset of barren
plateaus~\cite{Ragone:2023qbn,Fontana:2023mgx}.
Kazi et
al.~\cite{kazi2024analyzing} use Pauli DLA classifications to analyze
multi-angle MaxCut-QAOA on connected graphs.  For $G=(V,E)$, the
generator set is
\begin{equation*}
  \mathcal{G}_{\mathrm{free}}
  = \{X_v : v\in V\}\cup\{Z_uZ_v : \{u,v\}\in E\},
\end{equation*}
and
$\mathfrak{g}_{\mathrm{free}}
=\langle i\mathcal{G}_{\mathrm{free}} \rangle_{\operatorname{Lie}} $.

\begin{figure*}[htb!]
  \centering
  \includegraphics[width=0.6\linewidth]{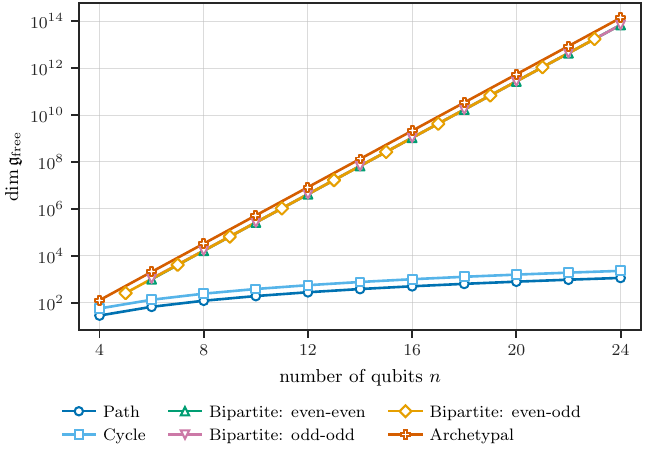}
  \caption{Analytic multi-angle MaxCut-QAOA DLA dimensions from Kazi
  et al. (solid curves) and PauLie classifications (open markers) for
  representative graphs from the path, cycle, bipartite parity, and
  archetypal classes.}
  \label{fig:kazi-comparison}
\end{figure*}

For this multi-angle ansatz, \cite[Theorem 1]{kazi2024analyzing} classifies the
DLA for every connected graph with $n\geq2$ vertices. The six cases are:
paths; cycles; connected bipartite graphs that are neither paths nor
cycles, split according to whether the bipartition sizes are
even-even, even-odd, or odd-odd; and archetypal graphs, i.e., connected
graphs that are neither bipartite nor cycles. 
For representatives of each of the six graph families at increasing $n$, PauLie returns the predicted algebra and dimension, as shown in Figure~\ref{fig:kazi-comparison}.

\cite[Corollary 2]{kazi2024analyzing}
 connects the DLA
classification to trainability for
archetypal graphs with $n>3$.  If a sufficiently deep multi-angle QAOA
circuit induces an approximate unitary $2$-design, then, with $d=2^n$,
\begin{align}
 \operatorname{Var}_{\boldsymbol\theta}
 [\partial_{\vartheta}C(\boldsymbol\theta)]
 &=\frac{4d^2|E|}{(d^2-4)(d+2)}
 \leq \frac{4n^2}{2^n}.
 \label{eq:kazi-gradient-variance}
\end{align}
Gradients thus vanish exponentially in $n$: the barren-plateau
phenomenon. 

The comparison is confined to the multi-angle variant, whose generators are
individual Pauli strings. With shared angles each generator becomes a linear
combination of Pauli strings; the generated algebra is then contained in the
one generated by the individual terms, but in general strictly smaller and
coefficient-dependent, so PauLie's classification does not apply directly.
Mao et al.~\cite{mao2025qaoamaxcut} treat that regime for MaxCut and find a
DLA dimension of $\Theta(4^n)$ for almost all graphs.

\section{Discussion and Future Work}\label{sec:conclusion}

PauLie brings DLA classification into the range of problem 
sizes relevant to current quantum algorithm research. Systems 
that previously required case-by-case analysis can now be classified 
routinely, which we expect to change how DLA structure enters 
algorithm design.

The framework is well suited to act as a routing oracle upstream 
of heavier methods. Exact Cartan decomposition 
\cite{wierichs2025recursivecartandecompositionsunitary}, 
classical simulation of polynomial-DLA systems \cite{Goh_2025}, 
and barren-plateau analysis of variational ansätze 
\cite{kazi2024analyzing} all depend on the DLA. PauLie determines the DLA efficiently, ensuring that approximate methods are applied only 
when necessary. The same capability supports numerical 
investigation of questions previously accessible only through 
manual analysis, such as generator-set optimality \cite{Smith_2025} 
and the construction of generator sets whose DLA is a direct power 
$\mathfrak{g}^{\oplus k}$ of a given algebra, using only 
logarithmically many extra qubits 
\cite{allcock2025generatingdirectpowersdynamical}.

Several extensions follow naturally. Since the canonical graphs obtained by PauLie are perfect, they are in particular \(\hbar\)-perfect in the sense of \cite{xu2025simultaneousvariancespaulistrings}. Consequently, the full set of applications developed there applies directly to the canonical graphs: efficient entanglement-detection schemes, connections to the complexity of shadow tomography, tight uncertainty relations, and constructions of strong lower bounds on ground-state energies.
In particular, the Pauli strings labelling the
vertices of a canonical graph span the operator subspace in which 
corresponding entanglement-witness operators can be constructed.
Perfection also certifies tractability in magic detection. For Pauli
measurement sets whose anticommutation graph is perfect and free of active sign dependencies, Liu et al.~\cite{liu2026nonstabilizerness} show that the reduced stabilizer polytope
becomes efficiently solvable, with witness capacity bounded by the clique
number. The canonical generator sets are thus natural candidates for scalable nonstabilizerness witnesses.
For all of these applications, an important open question is to what extent they can be pulled back from the canonical representative to the original instance.

Further extensions concern the class of supported inputs. Termwise Hamiltonian encodings in the sense of \cite{Cubitt_2018} preserve the DLA isomorphism. Since the encoded image of a Pauli string may have a nontrivial Pauli expansion, this provides
a principled route towards DLA classifications for generators that are linear
combinations of Pauli strings. 
Beyond qubits, the graph-theoretic backend may also be extended to
generalized Pauli observables through the qudit frustration-graph
formalism of \cite{makuta2025frustrationgraphformalismqudit}.

Lastly, the implementation itself can be extended. 
PauLie returns the isomorphism type of the DLA rather than an explicit
basis, so applications that consume a basis still require closure.
Producing bases in the cases where the algebra is small enough for one to
be useful is a natural next step. Also,
building the anticommutation graph requires $\binom{|\mathcal{G}|}{2}$
pairwise commutation checks, each a parity computation on the bit-packed
representation, and all of them are independent of one another. Li et
al.~\cite{li2026paulialgebra} implement precisely this kind of bitwise Pauli arithmetic on GPUs. Their framework could therefore take the place of PauLie's backend and evaluate the checks in parallel.

\section*{Acknowledgments}
The authors thank the Unitary Foundation for supporting the package through two microgrants and mentorship. We are grateful to Hendrik Poulsen Nautrup for suggesting the framework's application to discovering optimal universal generator sets. We further acknowledge Tarin Teacharsripaitoon for assistance with code quality improvements, as well as the open-source community for their contributions.
\printbibliography
\end{document}